\documentclass[journal]{IEEEtran}

\usepackage{amssymb}
\usepackage{hyperref}

\newtheorem{theorem}{Theorem}

\newtheorem{lemma}[theorem]{Lemma}

\newtheorem{corollary}[theorem]{Corollary}

\newcommand{\R}{{\mathbb{R}}}
\newcommand{\N}{{\mathbb{N}}}
\newcommand{\HB}{H_{\rm Burg}}
\newcommand{\sgn}{{\rm sgn}}

\ifCLASSINFOpdf
\else
\fi

\begin{document}
\title{A generalization of majorization that characterizes Shannon entropy}
\author{Markus~P.~M\"uller
\thanks{M.\ P.\ M\"uller is with the Departments of Applied Mathematics and Philosophy, University of Western Ontario,
Middlesex College, 1151 Richmond Street, London, ON N6A 5B7, Canada, and the Perimeter Institute
for Theoretical Physics, 31 Caroline St N, Waterloo, ON N2L 2Y5, Canada; email: markus@mpmueller.net.}and Michele Pastena\thanks{M.\ Pastena is
with the Institute for Theoretical Physics, Universit\"at Heidelberg, Philosophenweg 19, 69120 Heidelberg, Germany. email: pastena@thphys.uni-heidelberg.de.}}

\maketitle

\begin{abstract}
We introduce a binary relation on the finite discrete probability distributions which generalizes
notions of majorization that have been studied in quantum information theory.
Motivated by questions in thermodynamics, our relation describes the transitions induced by bistochastic maps
in the presence of additional auxiliary systems which may become correlated in the process.
We show that this relation is completely characterized by Shannon entropy $H$, which yields an interpretation of $H$ in resource-theoretic terms, and admits a particularly simple proof of a known characterization of $H$ in terms of natural information-theoretic properties.
\end{abstract}

\IEEEpeerreviewmaketitle

\section{Introduction}
\IEEEPARstart{M}{ajorization} and its relation to entropy plays a crucial role in many areas of probability and
information theory~\cite{MarshallOlkin}. A discrete probability distribution $p=(p_1,\ldots,p_n)$ is said to \emph{majorize} another
probability distribution $q=(q_1,\ldots,q_n)$, denoted
\[
   p\succ q,
\]
if and only if there is a bistochastic\footnote{Bistochastic maps are assumed to be linear. That is, these are maps that are
represented by bistochastic matrices, i.e.\ matrices with non-negative entries such that columns and rows both add up to one.} map $\Phi$ such that $q=\Phi(p)$.
The bistochastic maps are exactly the convex combinations of
permutations; therefore, $q$ is a random mixture of reshufflings of $p$ and in this sense \emph{more disordered} than $p$.

Since disorder and entropy are recurrent themes in thermodynamics, it comes as no surprise that majorization plays a major role there as well.
In particular, bistochastic maps and the majorization relation have been shown to determine the thermodynamically allowed state transitions
of systems out of equilibrium in the absence of energy constraints~\cite{Ruch,FundamentalLimitations,Brandao}. Similarly, majorization has been shown
to determine the interconvertibility of entangled pure quantum states by local operations and classical communication~\cite{Nielsen,JonathanPlenio}.

Mathematically, these applications have led to the study of majorization in the context of joint distributions of several random variables, in particular product distributions~\cite{Klimesh,Turgut,Brandao}. These appear naturally in the context of \emph{resource theories}~\cite{Coecke}, where random variables represent physical systems, and one asks how certain allowed transformations (such as bistochastic maps) are able to interconvert a given state of a physical system into another one. The interplay of the states of \emph{several} physical systems is of obvious interest, with the intuition that sometimes the presence of one physical system (say, a battery) can help to perform state transitions on another physical system (say, a laser pointer).

In this paper, we introduce a multipartite notion of majorization which is meant to elucidate the relation between \emph{disorder and correlation}. In a nutshell, while majorization determines whether a transformation $p\to q$ is possible via bistochastic maps,
we study transformations of the form
\begin{equation}
   p\otimes \left(\strut r_1\otimes\ldots\otimes r_k\right)\to q\otimes r_{1,\ldots,k}
   \label{eqTransition}
\end{equation}
which map $p$ to $q$, but at the same time correlate $k$ auxiliary systems without changing their marginals. Here, $\otimes$ denotes the Kronecker product, i.e.\ $p\otimes r$ is a product distribution on two systems; $r_{1,\ldots,k}$ denotes a joint probability distribution on $k$ systems, with marginals $r_1,\ldots,r_k$. Given two distributions $p$ and $q$, we ask whether there exists some $k\in\N_0$ and $r_{1,\ldots,k}$ such that transition~(\ref{eqTransition}) is possible via some bistochastic map. We can also fix a given value of $k$, in which case~(\ref{eqTransition}) generalizes the notions of majorization $(k=0)$ and \emph{trumping}~\cite{Nielsen} $(k=1)$ that have been extensively studied in quantum information theory.

In the case where both $p$ and $q$ do not contain zeros, and are not identical up to permutation, we show in Theorem~\ref{TheMain} below that a transformation
of the form~(\ref{eqTransition}) is possible if and only if $H(p)<H(q)$, for $H$ the Shannon entropy.
Thus, the possibility or impossibility of transitions of the form~(\ref{eqTransition}) is completely characterized by Shannon entropy. Furthermore, this insight can be used to give a particularly simple proof of a version of a known characterization of Shannon entropy: Acz\'el et al.~\cite[Lemma 5]{Aczel} have shown that $H$ is the unique real function (up to additive and multiplicative constants) on the probability distributions without zeros which is \emph{symmetric, additive, and subadditive}. If we additionally assume \emph{continuity}, then Theorem~\ref{TheMain} yields this characterization of $H$ as a simple corollary, cf.\ Corollary~\ref{CorCharacterization}.

While the detailed thermodynamic interpretation of~(\ref{eqTransition}) has been discussed elsewhere~\cite{Lostaglio}, the main idea can be phrased in the language of resource theories: if $p\not\succ q$ such that $p$ cannot be transformed into $q$ by bistochastic maps, but nevertheless $H(p)<H(q)$ such that~(\ref{eqTransition}) is possible, then \emph{stochastic independence is used as a resource}~\cite{Coecke,Fritz} in the transition. In other words: the additional creation of correlations
$r_1\otimes \ldots\otimes r_k\to r_{1,\ldots,k}$ in the auxiliary systems enables the otherwise impossible transition $p\to q$. This is comparable to the situation in Landauer's principle~\cite{Faist}, where the erasure of one bit of information, $(\frac 1 2,\frac 1 2)\to (1,0)$, can be accomplished at the additional expense of energy.

This paper is organized as follows. In Section~\ref{SecMainResults}, we give precise mathematical definitions and formulations
of our results, Theorem~\ref{TheMain} and Corollary~\ref{CorCharacterization}. Furthermore, we explain how the results fit into
the context of previous research on majorization and characterizations of entropy, and explain some results and open problems
related to the value of $k$ in~(\ref{eqTransition}). In Section~\ref{SecProofTheMain}, we give a proof of Theorem~\ref{TheMain},
which is accomplished by construction of a suitable auxiliary distribution $r_{1,\ldots,k}$ (however, with several non-trivial twists).
Section~\ref{SecProofCharacterization} shows how Corollary~\ref{CorCharacterization} follows as a simple consequence.
We conclude with Section~\ref{SecConclusions}, where we argue that our new relation~(\ref{eqTransition}) may be a special case of a wide variety of
interesting generalizations of majorization, characterizing transitions under consumptions of different kinds of information-theoretic resources.

\section{Main results and their context}
\label{SecMainResults}
In this paper, we are only considering finite discrete probability distributions. That is, in what follows,
a probability distribution is a vector $p=(p_1,\ldots,p_n)\in\R^n$ for some $n\in\N$ with the property
that all $p_i\geq 0$ and $\sum_{i=1}^n p_i=1$.
If we have a bipartite probability distribution, i.e.\ a joint
distribution of two random variables $A$ and $B$, then we denote this distribution by $p_{AB}$,
and its marginals by $p_A$ resp.\ $p_B$. In the case of $k>2$ random variables, we also
use the notation $p_{1,\ldots,k}$ for the joint distribution, and $p_i$ for its marginal on
the $i$-th random variable, which should not be confused with the $i$-th entry of a vector $p$.
The largest fixed number of systems or random variables that we consider explicitly will be five, which
we denote by $A,B,C,D,E$.

Majorization is defined in the following way. If $p,q\in\R^m$ are probability distributions, then\footnote{If majorization is
defined for arbitrary vectors $p,q\in\R^m$, one has to add the additional constraint $\sum_{i=1}^m p^\downarrow_i = \sum_{i=1}^m q^\downarrow_i$.
Since we are only considering probability distributions here, this condition is automatically satisfied and does not have to be specified.}
\begin{equation}
   p\succ q \quad\Leftrightarrow\quad \sum_{i=1}^k p^\downarrow_i \geq \sum_{i=1}^k q^\downarrow_i \mbox{ for all }k=1,\ldots,m,
   \label{eqDefMaj}
\end{equation}
where $p^\downarrow=(p_1^\downarrow,\ldots,p_n^\downarrow)$ denotes the reordering of the entries of $p$ in descending order,
i.e.\ $p^\downarrow_i=p_{\pi(i)}$ for some permutation $\pi$ such that $p^\downarrow_1\geq p^\downarrow_2\geq\ldots \geq p^\downarrow_m$.
This is equivalent~\cite{MarshallOlkin} to the existence of a bistochastic map $\Phi$, i.e.\ a linear map on $\R^m$ with $\Phi(1,\ldots,1)^\top=(1,\ldots,1)^\top$,
mapping probability distributions to probability distributions, such that $\Phi(p)=q$. Maps $\Phi$ of this kind are represented by bistochastic
matrices, i.e.\ square matrices with non-negative entries and row and column sums equal to one.
Given any probability distribution $p\in\R^m$, we define the \emph{rank} of $p$ as the number of non-zero entries of $p$.
That is,
\[
   {\rm rank}(p):=\#\{i\,\,|\,\, p_i\neq 0\}.
\]
Furthermore, the \emph{Shannon entropy} of any probability distribution $p\in\R^m$ is defined as
\[
   H(p):=-\sum_{i=1}^m p_i\log p_i,
\]
where $0\log 0:=0$ by definition, and $\log$ denotes the natural logarithm, i.e.\ $\exp(\log x)=x$.

\begin{table*}
\centering
\caption{Different majorization-like relations arising as special cases of~(\ref{eqCTrumping}).}
\label{TableMaj}
\begin{tabular}{|c|c|c|c|}
\hline
Case in~(\ref{eqCTrumping}) & notation & name & complete set of monotones\\
\hline
\hline
$k=0$ & $\succ$ & majorization & partial sums $S_k(p):=\sum_{i=1}^k p^\downarrow_i$ \\
& & & $(k=1,\ldots,m-1)$ if $p\in\R^m$\\
\hline
$k=1$ & $\succ_T$ & trumping & R\'enyi entropies $H_\alpha$ $(\alpha\in\R\setminus\{0\})$ \\
& & & and Burg entropy $\HB$\\
\hline
$k=2$ & ? & -- & ? \\
\hline
$k=3$ & $\succ_c$ & c-trumping & Shannon entropy $H$ \\
& & & and Hartley entropy $H_0$\\
\hline
$k\geq 4$ & same as $k=3$ & '' & '' \\
\hline
\end{tabular}
\end{table*}

With this notation at hand, we are ready to state our main result:
\begin{theorem}
\label{TheMain}
Let $p,q\in\R^m$ be probability distributions with $p^\downarrow\neq q^\downarrow$. Then there exists $k\in\N_0$ and a $k$-partite
probability distribution $r_{1,2,\ldots,k}$ such that
\begin{equation}
   p\otimes \left(r_1\otimes r_2\otimes\ldots\otimes r_k\right)\succ q\otimes r_{1,2\ldots,k}
   \label{eqCTrumping}
\end{equation}
if and only if ${\rm rank}(p)\leq {\rm rank}(q)$ and $H(p)<H(q)$. Moreover, we can always choose $k=3$.
\end{theorem}

Note that if $p^\downarrow =q^\downarrow$, then $q$ is a permutation of $p$, so $p\succ q$, and~(\ref{eqCTrumping})
is trivially true (with $k=0$). If $H(p)=H(q)$ and $p^\downarrow\neq q^ \downarrow$, then, strictly speaking, a transition of the form~(\ref{eqTransition}) is impossible. In this case, however, one can find full-rank approximations $q'$ that are arbitrarily close to $q$ and that satisfy $H(q')>H(q)=H(p)$, such that~(\ref{eqCTrumping}) holds for $q$ replaced by $q'$, allowing to obtain $q$ to arbitrary accuracy from $p$ via transitions of the form~(\ref{eqTransition}).

We now discuss the special cases of~(\ref{eqCTrumping}) for different values of $k$, summarized also in Table~\ref{TableMaj}.

If $k=0$, then~(\ref{eqCTrumping}) reduces to majorization itself. If we demand that~(\ref{eqCTrumping}) holds for $k=1$, we
ask for some distribution $r$ such that
\begin{equation}
   p\otimes r \succ q\otimes r.
   \label{eqTrumping}
\end{equation}
This notion has been introduced in entanglement theory~\cite{JonathanPlenio} and is called \emph{trumping}.
That is, \emph{$p$ trumps $q$}, denoted $p\succ_T q$, if and only if there is some distribution $r$ such that~(\ref{eqTrumping}) holds.
If $p\not \succ q$ but $p\succ_T q$ then the auxiliary distribution $r$ acts like a ``catalyst''.
The interpretation is similar to a catalyst in chemistry:
it enables transitions $p\to q$ that are impossible without its presence, but it is not consumed and can be reused after the process.

Motivated by this nomenclature, we call our new relation \emph{correlated trumping}, or \emph{c-trumping} and say that \emph{$p$ c-trumps $q$}, denoted $p\succ_c q$, if and only if there exists $k\in\N_0$ and $r_{1,2,\ldots,k}$ such that~(\ref{eqCTrumping}) holds. As stated in Theorem~\ref{TheMain}, the case $k=3$ is equivalent to leaving $k$ arbitrary, i.e.\ equivalent to c-trumping, and
so is any fixed value $k\geq 4$.

Understanding the case $k=2$ remains an interesting open problem. We conjecture that $k=2$
is equivalent to c-trumping, too, but have not been able to prove this.\footnote{We currently need $k=3$ catalysts in the proof of Theorem~\ref{TheMain} for the following reason: since R\'enyi entropies $H_\alpha$ with $0<\alpha<1$ behave very differently from those with
$1<\alpha<\infty$, the auxiliary distribution $r_{1,\ldots,k}$ is constructed in two steps, yielding a tripartite distribution.} An example of c-trumping with $k=2$ auxiliary systems can be found in~\cite{Lostaglio}, though in a more general framework in which systems are allowed to carry Hamiltonians (energy). Using the construction of Theorem 3 in the Supplemental Material of~\cite{Lostaglio}, one can obtain a pair of (high-dimensional) probability distributions $p,q$ from that example, such that $p\not\succ_T q$, but $p\otimes r_1\otimes r_2\succ q\otimes r_{12}$ for a suitable auxiliary distribution $r_{12}$, and thus $p\succ_c q$. While $k=2$ is sufficient for this particular choice of $p$ and $q$, we do not know whether it is in all cases.

For any two given distributions $p,q\in\R^m$, one can check directly whether $p\succ q$ by using the definition of majorization, (\ref{eqDefMaj}).
In contrast, the trumping relation $p\succ_T q$ is defined implicitly via the existence of a catalyst $r$ satisfying~(\ref{eqTrumping}) which cannot be checked directly.
Thus, it has been an open problem for some time to give necessary and sufficient conditions that allow one to decide whether or not $p\succ_T q$ holds.

This problem has been settled in the works of Klimesh~\cite{Klimesh} and Turgut~\cite{Turgut}. To understand their criterion, we need to define the R\'enyi and
Burg entropies which will play a major role later on in the proofs as well. For probability distributions $p\in\R^m$ and real parameters $\alpha\in\R\setminus\{0,1\}$,
we define the \emph{R\'enyi entropy of order $\alpha$} as
\[
   H_\alpha(p):=\frac{{\rm sgn}(\alpha)}{1-\alpha}\log\sum_{i=1}^m p_i^\alpha\qquad(\alpha\in\R\setminus\{0,1\}).
\]
Furthermore, we set
\begin{eqnarray*}
    H_\infty(p)&:=&-\log\max_i p_i,\qquad H_{-\infty}(p):=\log\min_i p_i,\\
    H_1(p)&:=&H(p),\qquad H_0(p):=\log {\rm rank}(p).
\end{eqnarray*}
This choice of definition ensures continuity of $H_\alpha$ in $\alpha$ except at $\alpha=0$, in the sense that
\begin{eqnarray*}
   \lim_{\alpha\to\infty}H_\alpha(p)=H_\infty(p),\qquad \lim_{\alpha\to 1}H_\alpha(p)=H_1(p),\\
   \lim_{\alpha\to -\infty}H_\alpha(p)=H_{-\infty}(p),\qquad
   \lim_{\alpha\searrow 0} H_\alpha(p)=H_0(p).
\end{eqnarray*}
However, $\lim_{\alpha\nearrow 0} H_\alpha(p)$ exists only if $p$ has ``full rank'', i.e.\ ${\rm rank}(p)=m$, in which case it equals $-\log m=-H_0(p)$.
The Burg entropy~\cite{Burg} is defined as
\[
   \HB(p):=\frac 1 m\sum_{i=1}^m \log p_i.
\]
Sometimes different conventions are used in the literature~\cite{Gour}; the prefactor $1/m$ ensures that $\HB$ is additive,
i.e.\ $\HB(p\otimes q)=\HB(p)+\HB(q)$.
Note that $\HB$ and $H_\alpha$ for $\alpha<0$ attain the value $-\infty$ if $p$ contains any zeros.
$H_0$ is also known as \emph{Hartley entropy} or \emph{max entropy}.

These entropies characterize the trumping relation as follows.
\begin{lemma}[Trumping~\cite{Klimesh,Turgut}]
\label{LemTrumping}
Let $p,q\in\R^m$ be probability distributions such that $p^\downarrow \neq q^\downarrow$, and such that at least one of them has full rank. Then $p\succ_T q$ if and only if
\begin{eqnarray*}
   H_\alpha(p)&<& H_\alpha(q) \mbox{ for all }\alpha\in\R\setminus\{0\},\mbox{ and}\\
   \HB(p)&<&\HB(q).
\end{eqnarray*}
\end{lemma}
Thus, fixing different values of $k$ in~(\ref{eqCTrumping}) naturally gives rise to different notions of entropy that
characterize the corresponding relations. A summary is shown in Table~\ref{TableMaj}.
Given some relation $\succ'$ on the probability distributions,
we say that a real function $S$ is a \emph{monotone} if $p\succ' q\Rightarrow S(p)\leq S(q)$. A set of monotones $(S_i)_{i\in I}$ will be called \emph{complete}
for the relation $\succ'$
if $S_i(p)<S_i(q)$ for all $i\in I$ implies that $p\succ' q$ (whether one would like to have strict or rather non-strict inequality, $S_i(p)\leq S_i(q)$, may depend on the context,
and does so in Table~\ref{TableMaj}). Thus, Lemma~\ref{LemTrumping} can be understood as saying that the R\'enyi and Burg entropies constitute a complete
set of monotones for the trumping relation. Similarly, Theorem~\ref{TheMain} says that the Shannon and Hartley entropies are a complete set of monotones
for c-trumping.

Our second result is an immediate consequence of Theorem~\ref{TheMain}. As mentioned above, while the result in Corollary~\ref{CorCharacterization} is not new (a slightly stronger version has been proved in~\cite{Aczel}), our proof seems to be considerably simpler once Theorem~\ref{TheMain} is established. Denote the probability distributions without zeros by
\[
   \Delta_n^+:=\left\{(p_1,\ldots,p_n)\in\mathbb{R}^n\,\,\left|\ p_i>0, \sum_{i=1}^n p_i=1\right.\right\},
\]
and set $\Delta^+:=\bigcup_{n\in\mathbb{N}} \Delta_n^+$. Then we have the following:
\begin{corollary}
\label{CorCharacterization}
A continuous function $S:\Delta^+\to\R$ satisfies the following three properties
\begin{itemize}
\item[(i)] \textbf{symmetry:} if $p,q\in\Delta^+$ are such that $p_i=q_{\pi(i)}$ for some permutation $\pi$ and all $i$, then $S(p)=S(q)$;
\item[(ii)] \textbf{subadditivity:} $S(p_{AB})\leq S(p_A\otimes p_B)$ for every bipartite probability distribution $p_{AB}\in\Delta^+$
with marginals $p_A$ and $p_B$;
\item[(iii)] \textbf{additivity:} $S(p_A\otimes p_B)=S(p_A)+S(p_B)$ for all $p_A,p_B\in\Delta^+$
\end{itemize}
if and only if it is of the form
\begin{equation}
   S(p)=c\cdot H(p)+c_n\mbox{ for all }p\in\Delta_n^+,\enspace n\in\N,
   \label{eqForm}
\end{equation}
where $H(p)=-\sum_i p_i\log p_i$ is Shannon entropy,
$c\geq 0$ some constant, and $c_n\in\R$ is some dimension-dependent constant with $c_{mn}=c_m+c_n$.
\end{corollary}

There is a vast literature on characterizations of Shannon entropy, see e.g.~\cite{AczelDaroczy,Csiszar,Chakrabarti,Baez}.
Our result is a slightly weaker version of the characterization in~\cite[Lemma 5]{Aczel}, which does not presuppose continuity, and (in addition to symmetry and additivity) only assumes \emph{weak subadditivity}, that is (ii) in the special case that $B$ has dimension two. It turns out that Theorem~\ref{TheMain} admits a straightforward proof of yet another version of Corollary~\ref{CorCharacterization}, which characterizes functions of the form~(\ref{eqForm}) as those that satisfy \emph{Schur concavity}, additivity, and subadditivity on $\Delta^+$, without assuming continuity. Schur concavity of $S$ means that $q=\Phi(p)$ for some bistochastic map $\Phi$ implies $S(q)\geq S(p)$, which is a property that one would intuitively expect from any ``measure of disorder''. However, since the proof is somewhat more involved than that of Corollary~\ref{CorCharacterization}, and since the result follows directly from those in~\cite{Aczel}, we omit the details.

\section{Proof of Theorem~\ref{TheMain}}
\label{SecProofTheMain}
We start by fixing some notation.
We say that a function $f:I\to\R$ with $I\subset\R$ is \emph{increasing} if $x< y\Rightarrow f(x)\leq f(y)$ for all $x,y\in I$, and that it is \emph{strictly increasing}
if $x< y\Rightarrow f(x)< f(y)$ (analogous definitions apply to \emph{decreasing / strictly decreasing}).
We will use the elementary limit identity~\cite{Brandao}
\begin{eqnarray}
   \HB(p)+\log m &=&  \lim_{\alpha\searrow 0} \frac {1-\alpha}\alpha \left(\strut H_\alpha(p)-\log m\right) \nonumber   \\
      &=&\lim_{\alpha\nearrow 0}\frac{1-\alpha}\alpha\left(-H_\alpha(p)-\log m\right).\label{eqBurgLimit}
\end{eqnarray}
Furthermore, note that R\'enyi entropy satisfies
\begin{equation}
   H_\alpha(p)\in \left\{
      \begin{array}{cl}
          \left[0, \log m\right] & \mbox{if }\alpha\geq 0 \\
         \left[-\infty,-\log m\right] & \mbox{if }\alpha <0,
      \end{array}
   \right.
   \label{eqRenyiIneq}
\end{equation}
and for every $\alpha\neq 0$, the maximal value $\sgn(\alpha) \log m$ is attained if and only if $p=\left(\frac 1 m,\ldots, \frac 1 m\right)$, cf.~\cite{Harremoes}.
The corresponding statement for the Burg entropy is $\HB(p)\leq -\log m$, with equality if and only if $p=\left(\frac 1 m,\ldots, \frac 1 m\right)$.

In the following, we will deal with multipartite (mostly bipartite) probability distributions. In the bipartite case,
we use the following notation. We denote the first system by $A$ (of size $m\in\N$), and the second by $B$ (of size $n\in\N$). Joint distributions
on $AB$ will be denoted as matrices with entries $(p_{AB})_{i,j}:=p(a=i,b=j)$. For example, if $p=p_A=(p_1,\ldots,p_m)$ and $q=q_B=(q_1,\ldots,q_n)$, then
\[
   p_A\otimes q_B=\left(
      \begin{array}{ccccc}
         p_1 q_1 & p_1 q_2 & p_1 q_3 & \ldots & p_1 q_n \\
         p_2 q_1 & p_2 q_2 & p_2 q_3 & \ldots & p_2 q_n  \\
         \vdots & \vdots & \vdots  & &\vdots \\
         p_m q_1 & p_m q_2 & p_m q_3 & \ldots & p_m q_n
      \end{array}
   \right)
\]
In general, the marginal distributions on $A$ resp.\ $B$ can be obtained by summing over the rows resp.\ columns of $p_{AB}$.
There is a specific family of bipartite probability distributions that will be important in what follows. If we have any probability distribution $q\equiv q_A=(q_1,\ldots,q_m)\in\R^m$, we
consider the specific extension
\begin{equation}
   q_{AB}:=\left(
      \begin{array}{ccccc}
         q_1-a_1 & \frac {a_1} n & \frac {a_1} n & \ldots & \frac{a_1} n \\
         q_2-a_2 & \frac {a_2} n & \frac {a_2} n & \ldots & \frac{a_2} n  \\
         \vdots & \vdots & \vdots  & &\vdots \\
         q_m-a_m & \frac {a_m} n & \frac {a_m} n & \ldots & \frac{a_m} n 
      \end{array}\right)
   \label{eqQAB}
\end{equation}
for any choice of $a_i\in[0,q_i]$ and $n\in\N$.
This is an $m\times(n+1)$ matrix, and a bipartite probability distribution with marginal $q_A$ on $A$ (which is what the word ``extension'' means here). Clearly
\[
   q_B=\left(1-a,\frac a n,\ldots,\frac a n\right)\in\R^{n+1},\qquad \mbox{where }a=\sum_{i=1}^m a_i.
\]
We need two lemmas. The first one is as follows.
\begin{lemma}
\label{LemPart1}
Let $p,q\in\R^m$ be probability distributions such that $q$ has full rank, $H(p)<H(q)$, and $q\neq \left(\frac 1 m,\ldots,\frac 1 m\right)$.
Then there exists some $\delta\in\left(0,\min_i q_i\right)$ and $N\in\N$ such that for $a_i:=q_i-\delta$ and
$q_{AB}$ as in~(\ref{eqQAB}), the following statement is true for all $n\geq N$:
\[
   H_\alpha(p_A\otimes q_B)<H_\alpha(q_{AB})\qquad\mbox{for all }\alpha\in[1,+\infty].
\]
\end{lemma}
\begin{IEEEproof}
Note that $p \neq\left(\frac 1 m,\ldots,\frac 1 m\right)$ because $H(p)<H(q)<\log m$.
In the following, we will always assume that $\alpha>1$, $\alpha\in\R$ (unless stated otherwise). With the given choice of $a_i$, we get $a=\sum_{i=1}^m a_i=1-m\delta$.
Consider the following expression:
\begin{eqnarray*}
   \Delta_n^{(\alpha)}&:=&H_\alpha(q_{AB})-H_\alpha(q_B)-H_\alpha(p_A)\\
   &=&\frac 1 {1-\alpha}\log
   \frac{m\delta^\alpha+n^{1-\alpha}\sum_{i=1}^m(q_i-\delta)^\alpha}{\left(\sum_{i=1}^m p_i^\alpha\right)\left(m^\alpha\delta^\alpha + (1-m\delta)^\alpha n^{1-\alpha}\right)}.
\end{eqnarray*}
We use the expression on the right-hand side to define $\Delta_n^{(\alpha)}$ also for non-integer $n\geq 1$.
We have to show that this expression is positive for all $\alpha$ if $n$ is large enough. In fact, in the limit,
\begin{equation}
   \lim_{n\to\infty}\Delta_n^{(\alpha)} = \log m -H_\alpha(p)>0\qquad \mbox{for all }\alpha>1,
   \label{eqLimit}
\end{equation}
which is however only a pointwise statement. We furthermore need the fact that
\begin{equation}
   \Delta_n^{(\alpha)}\mbox{ is strictly increasing in }n\mbox{ if }\alpha\in(1,\infty).
   \label{eqNonDecreasing}
\end{equation}
We prove this by checking that $\left(\sum_{i=1}^m p_i^\alpha\right)\exp\left((1-\alpha)\Delta_n^{(\alpha)}\right)$ is strictly increasing in $n^{1-\alpha}$. This expression is of the form $f(x):=(a+bx)/(c+dx)$ for $x:=n^{1-\alpha}$, where $a=m\delta^\alpha$, $b=\sum_{i=1}^m (q_i-\delta)^\alpha$, $c=m^\alpha\delta^\alpha$, and $d=(1-m\delta)^\alpha$. We have $f'(x)>0$ if and only if $ad<bc$, which (after some simplification) is equivalent to $H_\alpha\left(\left(\frac{q_i-\delta}{1-m\delta}\right)_i\right)< \log m$, and this inequality is satisfied since $q$ is not the uniform distribution and because of~(\ref{eqRenyiIneq}), proving~(\ref{eqNonDecreasing}).

Furthermore, for $\alpha=1$, we have
\begin{eqnarray*}
   \Delta_n^{(1)}&:=&H(q_{AB})-H(q_B)-H(p_A)\\ &=& m\delta\log m -\sum_{i=1}^m (q_i-\delta)\log\frac{q_i-\delta}{1-m\delta}-H(p_A),
\end{eqnarray*}
and this expression is independent of $n$. Since $\lim_{\delta\searrow 0}\Delta_n^{(1)}=H(q_A)-H(p_A)>0$, there exists some $\delta\in(0,\min_i q_i)$
such that with this choice of $\delta$, we have $\Delta_n^{(1)}>0$. So let us choose and fix this $\delta$ for all that follows. By continuity, for $n=1$,
there exists some $\varepsilon>0$ such that $\Delta_{n=1}^{(\alpha)}>0$ for all $1\leq\alpha\leq 1+\varepsilon$, and due to~(\ref{eqNonDecreasing})
\begin{equation}
   \Delta_n^{(\alpha)}>0\qquad\mbox{for all }n\in\N\mbox{ and }1\leq\alpha\leq 1+\varepsilon.
   \label{eqDeltaN1}
\end{equation}
Furthermore, if $n$ is large enough, then we have the exact equality
\begin{eqnarray*}
   \Delta_n^{(\infty)}&:=&H_\infty(q_{AB})-H_\infty(q_B)-H_\infty(p_A)\\
   &=&\log m -H_\infty(p)>0.
\end{eqnarray*}
Applying Lemma~\ref{LemDini} below to the family of functions $\alpha\mapsto\Delta_n^{(\alpha)}$ on the interval $[1+\varepsilon,\infty]$ (while taking into account~(\ref{eqLimit}) and~(\ref{eqNonDecreasing})) shows that there exists some $N\in\N$ such
that for all $n\geq N$, we have $\Delta_n^{(\alpha)}>0$ for all $\alpha$ in that interval. Together with~(\ref{eqDeltaN1}), this proves the claim.
\end{IEEEproof}

The second lemma which now follows is interesting in its own right. It gives a partial answer to the question under which conditions we can have a different
kind of ``correlated trumping relation'': instead of asking whether a transformation $p_A\otimes r_B\to q_A\otimes r_B$ is possible (corresponding to $\succ_T$),
one might allow that correlations between the two systems build up, such that $AB$ is finally described by a correlated distribution $q_{AB}$ with marginal $q_B=r_B$. In this sense, the ``catalyst'' would be retained in its original form, but correlated with the system that is to be transformed.

An example is given by the two distributions $p_{A}=\left(\frac{91}{100},\frac{1}{20},\frac{1}{25}\right)$ and
$q_{A}=\left(\frac{17}{20},\frac{7}{50},\frac{1}{100}\right)$. It is easy to see that $p_{A}\nsucc q_{A}$ (from the definition of majorization) and $p_{A}\nsucc_{T}q_{A}$
(since $H_{\alpha}\left(p_{A}\right)<H_{\alpha}\left(q_{A}\right)$
for all $\alpha\geq 1$ but not for all $\alpha<1$). However, if $q_{AB}$ is the correlated distribution in~(\ref{eqQAB}) with $n=1$ and $a_i=\frac 1 {120}$, then it turns out that $p_A\otimes q_B\succ_T q_{AB}$, as one can check by using Lemma~\ref{LemTrumping}. That is, there exists an additional system $C$ and a distribution $s_{C}$ such that $p_{A}\otimes\left(q_{B}\otimes s_{C}\right)\succ q_{AB}\otimes s_{C}$. If we denote the composite system $BC$ by $B'$ and set $q_{A B'}:=q_{AB}\otimes s_C$, then we have $p_A\otimes q_{B'}\succ q_{A B'}$. This example is a special case of the following result:

\begin{lemma}
\label{LemPart2}
Let $p,q\in\R^m$ be probability distributions such that $q$ has full rank, $q\neq\left(\frac 1 m,\ldots,\frac 1 m\right)$, and $H_\alpha(p)<H_\alpha(q)$ for all $\alpha\in[1,+\infty]$.
Then there exists some $a\in \left(0,m\cdot\min_i q_i\right)$  and $N\in\N$ such that for $q_{AB}$ as given in~(\ref{eqQAB}) with $a_i:=a/m$,
we have
\[
   p_A\otimes q_B\succ_T q_{AB}\qquad\mbox{for all }n\geq N.
\]
\end{lemma}
\begin{IEEEproof}
First consider the case that $p$ has full rank. Note that $p\neq\left(\frac 1 m,\ldots,\frac 1 m\right)$ since $H_1(p)<H_1(q)< \log m$.
We will use the criterion in Lemma~\ref{LemTrumping} to prove trumping. It holds
\begin{eqnarray*}
   \HB(q_{AB})=\frac 1 {m(n+1)}\sum_{i=1}^m \log\left(q_i-\frac a m\right)+\frac n {n+1}\log\frac a {mn},\\
   \HB(p_A\otimes q_B)=\frac 1 m\sum_{i=1}^m \log p_i 
   +\frac {\log(1-a)+n\log\frac a n}{n+1}.
\end{eqnarray*}
It is then elementary to see that the inequality $\HB(p_A\otimes q_B)<\HB(q_{AB})$ is equivalent to
\begin{eqnarray*}
   \frac 1 m\sum_{i=1}^m \log p_i+n\underbrace{\left(\frac 1 m \sum_{i=1}^m \log p_i+\log m\right)}_{(*)}+\log(1-a)\\
   <\frac 1 m\sum_{i=1}^m \log\left(q_i-\frac a m\right).
\end{eqnarray*}
Since $\frac 1 m\sum_{i=1}^m \log p_i=\HB(p)<-\log m$, the factor $(*)$ is negative. Hence this inequality is true if $n$ is large enough; in other words, there
exists $N(a)\in\N$ (which may depend on the choice of $a$) such that
\begin{equation}
   \HB(p_A\otimes q_B)<\HB(q_{AB})\qquad\mbox{for all }n\geq N(a).
   \label{eqRange1}
\end{equation}
For all $\alpha\in[-\infty,+\infty]$, define the quantity
\[
   \tilde\Delta_n^{(\alpha)}:=H_\alpha(q_{AB})-H_\alpha(q_B)-H_\alpha(p_A).
\]
If $\alpha=0$ this equals $0$; for general finite $\alpha\not\in\{ 0,1\}$, it is
\[
   \tilde\Delta_n^{(\alpha)}=\frac{\sgn(\alpha)}{1-\alpha}\log\frac{\sum_{i=1}^m \left(q_i-\frac a m\right)^\alpha+n^{1-\alpha}a^\alpha m^{1-\alpha}}
   {\left(\sum_{i=1}^m p_i^\alpha\right)\left(\strut (1-a)^\alpha+n^{1-\alpha}a^\alpha\right)}.
\]
First we prove the following:
\begin{equation}
   \tilde\Delta_n^{(\alpha)}\mbox{ is }\left\{
      \begin{array}{ll}
         \mbox{eventually constant in }n & \mbox{if }\alpha=-\infty \\
         \mbox{increasing in }n & \mbox{ if }-\infty<\alpha<1 \\
         \mbox{constant in }n & \mbox{ if }\alpha=1 \\
         \mbox{decreasing in }n & \mbox{ if }1<\alpha<+\infty \\
         \mbox{eventually constant in }n & \mbox{if }\alpha=+\infty.
      \end{array}
   \right.
   \label{eqTildeMonotonicity}
\end{equation}
By ``eventually constant'', we mean that there is some $N\in\N$ such that for all $n\geq N$, we have $\tilde\Delta_n^{(\alpha)}=\tilde\Delta_N^{(\alpha)}$.
This is the case for $\alpha=-\infty$ and $\alpha=+\infty$, because in this case, all entropies only depend on the minimal resp.\ maximal entries of
$q_{AB}$ resp.\ $q_B$; if $n$ is large, the location of these extrema is fixed, and direct calculation shows that all $n$-dependency cancels out.
The special case $\alpha=0$ is trivial; for $\alpha=1$, direct calculation shows that
\begin{eqnarray}
   \tilde\Delta_n^{(1)}&=&-\sum_{i=1}^m \left(q_i-\frac a m\right)\log\left(q_i-\frac a m\right)\nonumber\\
   &&+a\log m+(1-a)\log(1-a)-H(p)
   \label{eqShannon}
\end{eqnarray}
which is independent of $n$. For the remaining cases $\alpha\in\R\setminus\{0,1\}$, we check the monotonicity of $\left(\sum_{i=1}^m p_i^\alpha\right)\exp\left(\frac{1-\alpha}{{\rm sgn}(\alpha)}\tilde\Delta_n^{(\alpha)}\right)$ in $x:=n^{1-\alpha}$. This expression is of the form $f(x):=(a'+b'x)/(c'+d'x)$, with $a'=\sum_{i=1}^m \left(q_i-\frac a m\right)^\alpha$, $b'=a^\alpha m^{1-\alpha}$, $c'=(1-a)^\alpha$, and $d'=a^\alpha$. We have $f'(x)\geq 0$ if and only if $a'd'\leq b'c'$, which is equivalent to
\[
   \frac{1-\alpha}{{\rm sgn}(\alpha)}H_\alpha\left(\left(\frac{q_i-\frac a m}{1-a}\right)_i\right)\leq (1-\alpha)\log m.
\]
According to~(\ref{eqRenyiIneq}), this inequality is true for $0<\alpha<1$, but the inequality sign is reversed for $\alpha<0$ and $\alpha>1$. Taking care of the signs in all the different cases of $\alpha$ proves~(\ref{eqTildeMonotonicity}). By direct calculation, the large-$n$ limit of $\tilde\Delta_n^{(\alpha)}$ evaluates to
\begin{equation}
   \lim_{n\to\infty}\tilde\Delta_n^{(\alpha)}=\left\{
      \begin{array}{cl}
         -\log m-H_\alpha(p) & \mbox{if }\alpha\in[-\infty,0)\\
         \log m -H_\alpha(p) & \mbox{if }\alpha\in(0,1)\\
         \mbox{expression~(\ref{eqShannon}) above} & \mbox{if }\alpha=1\\
         H_\alpha\left(\left(\frac{q_i-a/m}{1-a}\right)_i\right)-H_\alpha(p) & \mbox{if }\alpha\in(1,+\infty]
      \end{array}
   \right.
   \label{eqAllLimits}
\end{equation}
which is discontinuous at $\alpha=0$ and $\alpha=1$.

So far, $a\in \left(0,m\cdot\min_i q_i\right)$ was arbitrary; now we are going to fix the value of $a$ in such a way that the limit in~(\ref{eqAllLimits}) is everywhere strictly positive. To this end, set
\[
   \tilde f_a(\alpha):=H_\alpha\left(\left(\frac{q_i-a/m}{1-a}\right)_i\right)-H_\alpha(p)\qquad(\alpha\in[1,+\infty]),
\]
and observe that this expression is decreasing in $a$ (for every fixed $\alpha\in[1,+\infty]$), as long as $a\in [0,m\cdot\min_i q_i]$. This follows from the fact that for $a,b$ in that interval with $a\leq b$, the probability distribution $[(q_i-b/m)/(1-b)]_i$ majorizes the
probability distribution $[(q_i-a/m)/(1-a)]_i$, and the R\'enyi entropies $H_\alpha$ with $\alpha\geq 1$ are Schur-concave~\cite{MarshallOlkin,Gour}.

Choose $j\in\N$ large enough such that $1/(j+1)<m\cdot\min_i q_i$, and for all $n\in\N$, set $f_n(\alpha):=\tilde f_{1/(n+j)}(\alpha)$. Then every $f_n$ is a continuous real function on $I:=[1,+\infty]$, and the monotonicity of $\tilde f_a$ in $a$ becomes $f_n(\alpha)\leq f_{n+1}(\alpha)$ for all $\alpha\in I$. Furthermore,
\[
   \lim_{n\to\infty} f_n(\alpha)=\lim_{a\searrow 0} \tilde f_a(\alpha)=
   H_\alpha(q)-H_\alpha(p)>0
\]
for all $\alpha\in I$. Thus, Lemma~\ref{LemDini} below proves that there is some $N\in\N$ such that $f_n(\alpha)>0$ for all $n\geq N$ and all $\alpha\in I$; in other words, there is some $a'\in (0,m\cdot\min_i q_i)$ such that $\tilde f_a(\alpha)>0$ for all $0\leq a \leq a'$ and all $\alpha\in I$.
Due to~(\ref{eqTildeMonotonicity}) and~(\ref{eqAllLimits}), we thus obtain
\[
   \tilde\Delta_n^{(\alpha)}\geq\lim_{n\to\infty}\tilde\Delta_n^{(\alpha)}=\tilde f_a(\alpha)>0
\]
for all $\alpha\in(1,+\infty]$, $a\in[0,a']$, and all $n\in\N$
(recall that $\tilde\Delta_n^{(\alpha)}$ depends on the choice of $a$). Due to~(\ref{eqShannon}), we have $\lim_{a\searrow 0}\tilde\Delta_n^{(1)}=H(q)-H(p)>0$,
so there exists $a\in(0,a')$ such that $\tilde\Delta_{n=1}^{(1)}>0$ for this choice of $a$.
\textbf{We now fix this value of $a$ for all that follows.}
Due to continuity, there exists $\varepsilon>0$ such that $\tilde\Delta_{n=1}^{(\alpha)}>0$ for all $\alpha\in[1-\varepsilon,1]$.
According to~(\ref{eqTildeMonotonicity}), this implies that $\tilde\Delta_n^{(\alpha)}>0$ for all $\alpha\in[1-\varepsilon,1]$ and all $n\in\N$.
In summary, we have achieved that
\begin{equation}
   \tilde\Delta_n^{(\alpha)}>0\qquad\mbox{for all }n\in\N,\enspace \alpha\in[1-\varepsilon,+\infty].
   \label{eqRange2}
\end{equation}
Next we consider $\alpha\in(0,1-\varepsilon)$. Since $\tilde\Delta_n^{(0)}=0$ for all $n$ is not useful as a special case, we define another quantity
\[
   \bar\Delta_n^{(\alpha)}:=\left\{
      \begin{array}{cl}
         \frac{1-\alpha}{|\alpha|} \tilde\Delta_n^{(\alpha)} & \mbox{if }\alpha\in\R\setminus\{0\} \\
         \strut \HB(q_{AB})-\HB(p_A\otimes q_B)
         & \mbox{if }\alpha=0.
      \end{array}
   \right.
\]
The resulting quantity is continuous in $\alpha$, also at $\alpha=0$ due to~(\ref{eqBurgLimit}).
Using that $\HB\left(\left(\frac{q_i-a/m}{1-a}\right)_i\right)<-\log m$, it is straightforward to check that
\[
   \frac\partial {\partial n}\bar\Delta_n^{(0)}=\frac{\log\frac{1-a}m - \frac 1 m\sum_{i=1}^m \log\left(q_i-\frac a m\right)}{(n+1)^2}> 0,
\]
hence $\bar\Delta_n^{(0)}$ is strictly increasing in $n$. The large-$n$ limit is
\[
   \lim_{n\to\infty} \bar\Delta_n^{(0)} = - \HB(p)-\log m>0
\]
since $p$ is not the uniform distribution.
Considering only $\alpha\in[0,1-\varepsilon]$, the $\bar\Delta_n^{(\alpha)}$ are an increasing sequence of continuous functions on this compact interval,
converging pointwise to a strictly positive continuous function due to~(\ref{eqTildeMonotonicity}), (\ref{eqAllLimits}), and~(\ref{eqBurgLimit}). Thus, Lemma~\ref{LemDini} below proves that there exists some
$N'\in\N$ such that $\bar\Delta_n^{(\alpha)}>0$ for all $n\geq N'$ and $\alpha\in[0,1-\varepsilon]$, hence
\begin{equation}
   \tilde\Delta_n^{(\alpha)}>0\qquad\mbox{for all }n\geq N',\enspace \alpha\in(0,1-\varepsilon].
   \label{eqRange3}
\end{equation}
Now we come to the case $\alpha<0$. According to~(\ref{eqTildeMonotonicity}) and~(\ref{eqAllLimits}), there exists $N''\in\N$ such that for all $n\geq N''$,
it holds $\tilde\Delta_n^{(-\infty)}=-\log m -H_{\-\infty}(p)>0$. Due to continuity, there is some $\alpha_-\in\R$ such that $\tilde\Delta_{N''}^{(\alpha)}>0$
for all $\alpha\in[-\infty,\alpha_-]$, and thus (again due to~(\ref{eqTildeMonotonicity}))
\begin{equation}
   \tilde\Delta_n^{(\alpha)}>0\qquad\mbox{for all }n\geq N'',\enspace \alpha\in[-\infty,\alpha_-].
   \label{eqRange4}
\end{equation}
Finally we treat the range $\alpha\in (\alpha_-,0)$. Arguing as above, the $\bar\Delta_n^{(\alpha)}$ are an increasing sequence of
continuous functions on the compact interval $[\alpha_-,0]$, converging pointwise to a strictly positive continuous function. According to Lemma~\ref{LemDini} below, there exists some $N'''\in\N$ such that $\bar\Delta_n^{(\alpha)}>0$ for all $n\geq N'''$, and thus
\begin{equation}
   \tilde\Delta_n^{(\alpha)}>0\qquad\mbox{for all }n\geq N''',\enspace \alpha\in[\alpha_-,0).
   \label{eqRange5}
\end{equation}
Combining~(\ref{eqRange1}), (\ref{eqRange2}), (\ref{eqRange3}), (\ref{eqRange4}), and (\ref{eqRange5}), and setting $N:=\max\{N(a),N',N'',N'''\}$, we get
\begin{eqnarray*}
   H_\alpha(p_A\otimes q_B)&<&H_\alpha(q_{AB})\mbox{ for all }\alpha\in\R\setminus\{0\},\mbox{ and}\\
   \HB(p_A\otimes q_B)&<&\HB(q_{AB})\qquad\mbox{for all }n\geq N.
\end{eqnarray*}
Clearly $(p_A\otimes q_B)^{\downarrow}\neq q_{AB}^{\downarrow}$, because otherwise we would have $H(p_A\otimes q_B)=H(q_{AB})$. Furthermore, $q_{AB}$ has full rank. Thus, Lemma~\ref{LemTrumping} proves that $p_A\otimes q_B\succ_T q_{AB}$.

We have proven the statement of the lemma in the case that $p$ has full rank. Now consider the case that ${\rm rank}(p)<m$. Since $q$ and thus $q_{AB}$ has full rank,
we only have to show that $H_{\alpha}(p_A\otimes q_B)<H_\alpha(q_{AB})$ for all $\alpha\in (0,+\infty)$. To this end, we can simply repeat the proof above with a few small
changes. First, the cases of Burg entropy and R\'enyi entropy for $\alpha<0$ can be ignored. Second, the proof of~(\ref{eqRange2}) remains valid, but the
proof of~(\ref{eqRange3}) has to be changed: instead of $\bar\Delta_n^{(\alpha)}$, we have to consider the quantity $\tilde\Delta_n^{(\alpha)}$ directly,
which now satisfies $\tilde\Delta_n^{(0)}=\log m -H_0(p)>0$ for all $n$. The rest of the argumentation remains unchanged, proving the statement of the
lemma also for the case that $p$ does not have full rank.
\end{IEEEproof}

The previous two lemmas have made use of the following basic result, which is a simple consequence of Dini's theorem.
\begin{lemma}
\label{LemDini}
Let $-\infty\leq a<b\leq+\infty$, and $(f_n)_{n\in\N}$ a family of continuous real functions on $I:=[a,b]$. (If $b=+\infty$ we demand that every $f_n$ is continuous on $[a,+\infty)$ and that the limit $f_n(+\infty):=\lim_{x\to +\infty} f_n(x)$ exists for all $n$; analogously for the case $a=-\infty$). Suppose that the family of functions is increasing, i.e.\ $f_n(x)\leq f_{n+1}(x)$ for all $x\in I$, and that $\lim_{n\to\infty}f_n(x)=f(x)$ for some continuous strictly positive function $f:I\to\R$. Then there is some $N\in\N$ such that $f_n(x)>0$ for all $n\geq N$ and all $x\in I$.
\end{lemma}
\begin{IEEEproof}
If either $a=-\infty$ or $b=+\infty$ (or both), we can consider the functions $\tilde f_n(y):=f_n(\tan y)$ for $y\in \arctan I=[\arctan a,\arctan b]\subset [-\pi/2,\pi/2]$ instead of the $f_n$, and in this way reduce everything to the case that $I\subset\R$. But in this case, Dini's theorem proves that the convergence $f_n\to f$ is uniform, hence with $\epsilon:=\min_{x\in I} f(x)>0$ there is some $N\in\N$ such that $|f(x)-f_n(x)|<\epsilon/2$ and therefore $f_n(x)>0$ for all $x\in I$ and $n\geq N$.
\end{IEEEproof}

Combining Lemmas~\ref{LemPart1} and~\ref{LemPart2} yields a first formulation of our main result.
\begin{lemma}
\label{LemAlmostMain}
Let $p,q\in\R^m$ be probability distributions such that $q$ has full rank. If $H(p)<H(q)$ then there exists $k\in\N$ (in fact, we can
always choose $k=3$) and a $k$-partite distribution $r_{1,2,\ldots,k}$ with marginals $r_1,r_2,\ldots, r_k$ such that
\[
   p\otimes \left(\strut r_1\otimes r_2\otimes\ldots\otimes r_k\right) \succ q\otimes r_{1,2,\ldots,k}.
\]
\end{lemma}
\begin{IEEEproof}
The special case that $q=\left(\frac 1 m,\ldots,\frac 1 m\right)$ is trivial: in this case $p\succ q$, and we can simply set $k=0$ (no auxiliary system), or alternatively $k=1$
with an arbitrary auxiliary distribution.

So suppose $q\neq\left(\frac 1 m,\ldots,\frac 1 m\right)$. We first apply Lemma~\ref{LemPart1} to conclude that there exists some extension $q_{AB}$ of $q=q_A$ such
$H_\alpha(p_A\otimes q_B)<H_\alpha(q_{AB})$ for all $\alpha\in[1,+\infty]$. Clearly the extension $q_{AB}$ given in that lemma has full rank, but is not a uniform distribution.
Therefore, we can apply Lemma~\ref{LemPart2} to the two distributions $p_A\otimes q_B$ and $q_{AB}$, and obtain the existence of an extension $q_{ABC}$
(introducing a third system $C$) of $q_{AB}$ such that
\[
   (p_A\otimes q_B)\otimes q_C \succ_T q_{ABC}.
\]
By definition of trumping, there is an additional system $D$ and a catalyst (probability distribution) $c_D$ on $D$ such that
\[
   p_A\otimes q_B \otimes q_C \otimes c_D \succ q_{ABC}\otimes c_D.
\]
Since the majorization relation is preserved under the tensor product with another probability distribution,
we obtain
\[
   p_A\otimes q_B\otimes q_C \otimes c_D \otimes q_E \succ q_{ABC}\otimes c_D \otimes q_E,
\]
where $q_E=q=q_A$ is another copy of $q$ (note however that $q_B$ and $q_C$ are in general \emph{not} copies of $q=q_A$).
Swapping systems $A$ and $E$ on the right-hand side does not alter the probability values and the majorization order, thus
\[
   p_A\otimes(q_E\otimes q_B\otimes q_C \otimes c_D)\succ q_A\otimes(q_{EBC}\otimes c_D).
\]
If we regard $CD$ as a single system (which we may, since the marginal of $q_{EBC}\otimes c_D$ on $CD$ is $q_C\otimes c_D$),
we see that we have $k=3$ subsystems in addition to system $A$.
\end{IEEEproof}

Now we are ready to prove our main result, Theorem~\ref{TheMain}.
\begin{IEEEproof}
Suppose there exists an auxiliary distribution $r_{1,2,\ldots,k}$ with the stated properties. Then we can apply additivity and subadditivity~\cite{Aczel,Linden}
as well as Schur concavity~\cite{MarshallOlkin} of the R\'enyi entropies of orders $\alpha=0$ and $\alpha=1$ (Hartley and Shannon entropy) and obtain
\begin{eqnarray*}
   H_\alpha(p)+\sum_{i=1}^k H_\alpha(r_i)&\leq& H_\alpha(q)+H_\alpha(r_{1,2,\ldots,k})\\
   &\leq& H_\alpha(q)+\sum_{i=1}^k H_\alpha(r_i).
\end{eqnarray*}
Since $H_0(p)=\log{\rm rank}(p)$, this shows that ${\rm rank}(p)\leq{\rm rank}(q)$. For Shannon entropy $H=H_1$, we obtain equality in the second
inequality of this expression (subadditivity) if and only if $r_{1,2,\ldots,k}=r_1\otimes r_2\otimes\ldots\otimes r_k$; this follows inductively from the fact
that the mutual information of two random variables is zero if and only if the joint bipartite probability distribution factorizes~\cite{CoverThomas}. So if we had $H(p)=H(q)$ then
$p\otimes\left(r_1\otimes r_2\otimes\ldots\otimes r_k\right)\succ q \otimes\left(r_1\otimes r_2\otimes\ldots\otimes r_k\right)$, or $p\succ_T q$.
But then Lemma~\ref{LemTrumping} (possibly after removing common zeros from $p$ and $q$ as in the following paragraph below) would prove that $H(p)<H(q)$, which is a contradiction.

Conversely, suppose that $p,q\in\R^m$ are probability distributions that are not equal up to permutation and satisfy ${\rm rank}(p)\leq{\rm rank}(q)$
and $H(p)<H(q)$. Without loss of generality we may assume that $p^{\downarrow}=p$ and $q^{\downarrow}=q$, i.e.\ that the entries of $p$ and $q$
are in descending order. Let $\ell:={\rm rank}(q)$, then $\ell\leq m$ and $q=\tilde q\oplus 0_{m-\ell}$, where $\tilde q =(q_1,\ldots,q_\ell)\in\R^\ell$
has full rank, and $0_{m-\ell}=(0,\ldots,0)\in\R^{m-\ell}$ is the zero vector of dimension $m-\ell$. Since ${\rm rank}(p)\leq {\rm rank}(q)=\ell$, we can
also write $p=\tilde p\oplus 0_{m-\ell}$, where $\tilde p\in\R^\ell$ does not necessarily have full rank. Then~(\ref{eqCTrumping}) for some probability
distribution $r_{1,2,\ldots,k}$ is equivalent to
\[
   \tilde p\otimes\left(r_1\otimes r_2\otimes\ldots\otimes r_k\right)\succ \tilde q\otimes r_{1,2,\ldots,k}.
\]
Since $H(\tilde p)=H(p)<H(q)=H(\tilde q)$, and since $\tilde q$ has full rank, Lemma~\ref{LemAlmostMain} applies and shows that a probability
distribution $r_{1,2,\ldots,k}$ exists that satisfies this relation.
\end{IEEEproof}
Similarly as for catalytic majorization~\cite{JonathanPlenio}, it is easy to show that auxiliary distributions $r_i$ which are either fully mixed
(i.e.\ equal to $\left(\frac 1 n,\ldots, \frac 1 n\right)\in\R^n$ for some $n$) or pure (i.e.\ contain only zeros and ones) are useless; they can be removed
without altering the c-trumping relation. In other words, we may assume that every auxiliary system $r_i\in\R^n$
appearing in~(\ref{eqCTrumping}) has Shannon entropy strictly positive and strictly less than $\log n$.

\section{Proof of Corollary~\ref{CorCharacterization}}
\label{SecProofCharacterization}
\begin{IEEEproof}
It is obvious that every function $S:\Delta^+\to\R$ of the form~(\ref{eqForm}) is continuous and has properties (i), (ii) and (iii).
It remains to show that converse; so suppose that $S$ is a continuous real function on $\Delta^+$ that has properties (i), (ii), and (iii).
Use the notation
\[
   \eta_n:=\left(\frac 1 n, \ldots, \frac 1 n\right)\in\Delta_n^+\qquad (n\in\N\setminus\{1\}),
\]
and define the ``negentropies'' for all $p\in\Delta_m^+$, $m\in\N$, as
\begin{eqnarray}
   I(p)&:=& H(\eta_m)-H(p) = \log m -H(p),\nonumber\\
   J(p)&:=& S(\eta_m)-S(p).
   \label{eqDefinition}
\end{eqnarray}
We claim that $J$ is non-negative. This can be seen from a simple argument which, for notational reasons, we give only for $m=3$, but which obviously works for all $m$. Using additivity, symmetry, and subadditivity (recalling our matrix notation for bipartite distributions), we obtain
\begin{eqnarray*}
	S(\eta_3)+S(p)&=& S(\eta_3\otimes p)=S\left(\begin{array}{ccc} p_1/3 & p_2/3 & p_3/3 \\ 	p_1/3 & p_2/3 & p_3/3 \\ p_1/3 & p_2/3 & p_3/3 \end{array}\right)\\
	&=&S\left(\begin{array}{ccc} p_1/3 & p_2/3 & p_3/3 \\ 	p_2/3 & p_3/3 & p_1/3 \\ p_3/3 & p_1/3 & p_2/3 \end{array}\right) \\
	&\leq& S(\eta_3)+S(\eta_3),
\end{eqnarray*}
hence $S(p)\leq S(\eta_3)$, and in general $S(p)\leq S(\eta_m)$ for all $p\in\Delta_m^+$ by the same argument.

We will now show that $S$ is Schur-concave. Suppose that $r,s\in\Delta^+$ satisfy $r\succ s$ (implying in particular that these distributions have the same number of entries). Then, for every $\epsilon>0$, there is a distribution $s_\epsilon$ with $\|s-s_\epsilon\|<\epsilon$ and a permutation $\pi_{AB}$ on a bipartite system $AB$ such that
\[
   s_\epsilon=\left[\pi_{AB}(r_A\otimes\eta_B)\right]_A.
\]
That is, $s$ can be obtained to arbitrary accuracy by bringing in an extra system $B$ in a uniform distribution, performing a suitable global permutation, and restricting to the marginal on $A$. This fact has been used extensively in quantum thermodynamics~\cite{HHO,Gour,Ruch}. Thus
\begin{eqnarray*}
	S(r_A)+S(\eta_B)&=& S(r_A\otimes \eta_B)
	=S\left(\strut \pi_{AB}(r_A\otimes\eta_B)\right)\\
	&\leq& S\left(\strut [\pi_{AB}(r_A\otimes\eta_B)]_A\right)+
	S\left(\strut [\ldots]_B\right)\\
	&\leq& S(s_\epsilon)+S(\eta_B).
\end{eqnarray*}
By continuity, it follows that $S(r)\leq S(s)$, that is, Schur-concavity.

We claim that for all $p,q\in\Delta^+$,
\begin{equation}
   I(p)\geq I(q) \Rightarrow J(p)\geq J(q).
   \label{eqMonotonicity1}
\end{equation}
To see this, suppose that $p\in\Delta_m^+$ and $q\in\Delta_n^+$ with $I(p)\geq I(q)$. If $q=\eta_n$ then $J(q)=0\leq J(p)$ as claimed. Otherwise, for every $\epsilon\in (0,1)$, define $q_\epsilon:=(1-\epsilon)q+\epsilon\eta_n$, then
\[
   H(p\otimes\eta_n)\leq H(q\otimes\eta_m)<H(q_\epsilon\otimes\eta_m).
\]
Thus, according to Theorem~\ref{TheMain}, for every $\epsilon\in(0,1)$ there exists some tripartite distribution $c_{123}$ such that
\[
   p\otimes \eta_n\otimes c_1\otimes c_2 \otimes c_3 \succ q_\epsilon\otimes \eta_m\otimes c_{123}.
\]
Using (ii), (iii), and Schur-concavity, we get
\begin{eqnarray*}
   S(p)&+&S(\eta_n)+S(c_1)+S(c_2)+S(c_3)\\
   &=& S(p\otimes \eta_n\otimes c_1\otimes c_2\otimes c_3)\\
   &\leq& S(q_\epsilon\otimes \eta_m\otimes c_{123})\\
   &=& S(q_\epsilon)+S(\eta_m)+S(c_{123})\\
   &\leq& S(q_\epsilon)+S(\eta_m)+S(c_1)+S(c_2)+S(c_3).
\end{eqnarray*}
Therefore $J(p)\geq J(q_\epsilon)$, and by continuity $J(p)\geq J(q)$. This proves~(\ref{eqMonotonicity1}). If $I(p)=I(q)$ then we have~(\ref{eqMonotonicity1}) in both directions, hence $J(p)=J(q)$. Thus there is a function $f:[0,\infty)\to\R$ with $f(0)=0$ such that $J(p)=f(I(p))$ for all $p\in\Delta^+$.
According to~(\ref{eqMonotonicity1}), this function $f$ is increasing. If $x,y\geq 0$, let $p,q\in\Delta^+$ be
distributions with $I(p)=x$ and $I(q)=y$, then
\begin{eqnarray*}
   f(x+y)&=& f\left(I(p)+I(q)\right)=f(I(p\otimes q)) = J(p\otimes q)\\
   &=& J(p)+J(q)=f(I(p))+f(I(q))\\
   &=&f(x)+f(y).
\end{eqnarray*}
Thus, $f$ is an additive monotone function, and it is well-known (and easy to check) that all functions of this kind are linear.
Hence there is a constant $c\in\R$ such that $J(p)=c\cdot I(p)$, and this constant cannot be negative due to~(\ref{eqMonotonicity1}).
Recalling the definition~(\ref{eqDefinition}), we get for $p\in\Delta_m^+$
\[
   S(p)=c\cdot H(p)+\underbrace{S(\eta_m)-c\log m}_{=:c_m}
\]
and from $\eta_{mn}=\eta_m\otimes\eta_n$ is is easy to check that $c_{mn}=c_m+c_n$.
\end{IEEEproof}

A few comments are in place regarding the statement of this corollary.
Note that
the additivity property $c_{mn}=c_m+c_n$ for the dimension-dependent constants does not automatically imply that $c_n=b\cdot \log n$
for some constant $b\in\R$. While this is a possible choice of $c_n$, there are other choices, and one needs additional assumptions
to conclude that $c_n$ is a logarithm, cf.~\cite{Mate}.

It is well-known that Hartley entropy $H_0$ is symmetric, additive, and subadditive. However, if $p\in\Delta_n^+$, i.e.\ $p$ does not contain zeros,
then $H_0(p)=\log n$, i.e.\ a dimension-dependent constant, which is covered by our theorem.

From the structure of the proof, one can conclude that the actual mathematically ``natural'' quantity is not Shannon entropy $H$ itself,
but negentropy $I(p):=\log n-H(p)$ (for $p\in\Delta_n^+$). This resembles the fact that $I$ (and not $H$) turns out to be the relevant
quantity to describe the amount of extractable work in many situations in thermodynamics, cf.~\cite{OHHH,Gour}.

Note that the R\'enyi entropies $H_\alpha$ and the Burg entropy $\HB$
are continuous, symmetric, and additive, and so are non-negative (discrete or continuous) linear combinations of them. It is therefore natural to conjecture that these are the only real functions on $\Delta^+$ that satisfy the analog of Corollary~\ref{CorCharacterization}
if the assumption of subadditivity (ii) is dropped. This conjecture resembles Example 7.10 in~\cite{Fritz}. However, it is not clear whether the methods of this
paper allow to contribute in any way to a resolution of this conjecture.

\section{Conclusions}
\label{SecConclusions}
We have introduced a new relation on the finite discrete probability distributions, called \emph{c-trumping}, which is part of a series of natural generalizations of the notions of majorization and trumping as studied in quantum information theory. It is meant to elucidate the relation between correlation and disorder, and turns out to be completely characterized by Shannon entropy $H$. We have also shown that this insight can be used to obtain a very simple proof of a weaker version of Acz\'el et al.'s characterization result~\cite{Aczel}.

It has been noted before that the notion of trumping, or catalysis, is very sensitive to the detailed requirements on how the catalysts are retained in the end. For example, if~(\ref{eqTrumping}) is replaced by the weaker condition that $p\otimes r \succ q\otimes r'$, where $r'$ is $\epsilon$-close in variation distance to $r$ for some fixed $\epsilon>0$, then \emph{all} transitions from any $p$ to any $q$ become possible, and the resulting relation becomes trivial. This phenomenon has been called \emph{embezzling} in the context of entanglement theory~\cite{vanDam} and thermodynamics~\cite{Brandao}. If one demands that the variation distance is smaller than $\epsilon$ \emph{divided by the logarithm of the catalyst dimension}, then it turns out that the Shannon entropy $H$ determines the allowed transitions~\cite{Brandao}, which is somewhat similar to our result.

So can Theorem~\ref{TheMain} be interpreted as an instance of embezzling? We do not think so. Note that we demand that the  auxiliary systems $r_1,\ldots,r_k$ preserve their local states exactly. More generally, while it has been argued in~\cite{Brandao} that ``closeness in variation distance'' is simply not a physically meaningful requirement, we think that ``local preservation of the auxiliary distributions'' is a physically well-motivated condition: restrictions on transformations in physics usually arise from conservation laws. But in most situations, conserved quantities (like energy or angular momentum) are sums of local quantities as long as interaction terms can be neglected. In this sense, our result says in what way we can exploit auxiliary systems as resources, if these systems are forced to preserve their local states due to local conservation laws.

If local states \emph{are} allowed to change, then physical intuition expects these systems to thermalize; in the context of majorization, this amounts to getting closer to the uniform distribution. This paper can be interpreted as studying the complementary situation in which local states are forced to be fixed. Theorem~\ref{TheMain} then gives a classification of what is possible in this regime, and suggests that there might be some situations of this kind in physics where correlations build up spontaneously.

The c-trumping relation represents a special instance of a more general problem: instead of asking whether a given distribution $p$ can be transformed into another distribution $q$ by some bistochastic map, we can ask whether this is possible if some additional resources are consumed or produced during the transformation.

More formally, think of some set of input auxiliary distributions $\mathcal{I}$, and to every $r\in\mathcal{I}$
a corresponding set of output distributions $\mathcal{O}_r$. We may then ask whether there exist auxiliary distributions
$r\in\mathcal{I}$ and $r'\in\mathcal{O}_r$ such that $p\otimes r \succ q\otimes r'$. If $r$ is in some sense ``more valuable'' than $r'$, then the transition $p\to q$ can be accomplished at the cost of some auxiliary resource; otherwise we have a resource yield. While this formulation represents a simplification
of the general idea of a resource theory~\cite{Coecke,Fritz}, it may already lead to non-trivial but mathematically tractable relations on probability distributions,
with in some cases interesting consequences for thermodynamics.

In the case of c-trumping, $\mathcal{I}$ is the set of product distributions, $\mathcal{O}_r$ is the set of multipartite distributions that have the same marginals as $r\in\mathcal{I}$, and transitions involve a cost of stochastic independence. A different example is given by the notion of \emph{lambda-majorization} that has been introduced in~\cite{Faist} to calculate the work cost of arbitrary processes such as Landauer erasure. They study transitions of the form
\[
   p\otimes \eta_2^{\otimes i}\otimes x_2^{\otimes (n-i)} \to  q\otimes \eta_2^{\otimes j}\otimes x_2^{\otimes(n-j)}
\]
via bistochastic maps, where $x_2=(1,0)$ is a ``pure bit'', and $\eta_2=\left(\frac 1 2,\frac 1 2\right)$. Given $p$ and $q$ of identical size, they ask for the maximal $\lambda:=i-j$ over all $n\in\N$ (arbitrary number of ``auxiliary bits'') such that a transition of this form is possible, i.e.\ the left-hand side majorizes the right-hand side. This is interpreted as extraction of work proportional to $\lambda$ by resorting to Landauer's principle. In our formalism, we can fix $\lambda\in\mathbb{Z}$, define $\mathcal{I}$ as the set of all distributions $r_{k,l}:=\eta_2^{\otimes k}\otimes x_2^{\otimes l}$ with arbitrary $k,l\in\N_0$, $k\geq\lambda$, and $\mathcal{O}_{k,l}$ as the set of all distributions of the form $\eta_2^{\otimes(k-\lambda)}\otimes x_2^{\otimes(l+\lambda)}$. This way, our formalism expresses the question whether work extraction proportional to the given value of $\lambda$ is possible.

As the results in this paper indicate, the study of generalized majorization relations of this kind may lead to surprising insights into the ``usefulness''
of information-theoretic properties. This contributes to the general question how different kinds of knowledge (represented by probability
distributions) can be ``put to work'' via interconversion, and in what way this is expressed by the values of entropy-like
quantities. Clearly, this kind of reasoning is not restricted to classical probability distributions, but can applied to quantum states as well,
which are the main subject of interest in quantum thermodynamics.

\section*{Acknowledgments}
We are grateful to Matteo Lostaglio, Jonathan Oppenheim, and Manfred Salmhofer for discussions. Furthermore, we would like to thank an anonymous referee for his or her thorough and insightful comments that allowed to simplify several proof steps and clarified the relation to~\cite{Aczel}. Research at Perimeter Institute is supported by the Government of Canada through Industry Canada and by the Province of Ontario through the Ministry of Research and Innovation. M.P.\ thanks the Heidelberg Graduate School of Fundamental Physics for financial support.

\ifCLASSOPTIONcaptionsoff
  \newpage
\fi


\begin{thebibliography}{99}
\bibitem{MarshallOlkin}
A.\ W.\ Marshall, I.\ Olkin, and B.\ C.\ Arnold, \emph{Inequalities: Theory of Majorization and Its Applications}, Springer, 2010.

\bibitem{Ruch}
E.\ Ruch and A.\ Mead, \emph{The Principle of Increasing Mixing Character and Some of Its Consequences}, Theor.\ Chim.\ Acta \textbf{41}, 95--117 (1976).

\bibitem{FundamentalLimitations}
M.\ Horodecki and J.\ Oppenheim, \emph{Fundamental limitations for quantum and nanoscale thermodynamics}, Nat.\ Comm.\ \textbf{4}, 2059 (2013).

\bibitem{Brandao}
F.\ G.\ S.\ L.\ Brand\~ao, M.\ Horodecki, N.\ H.\ Y.\ Ng, J.\ Oppenheim, and S.\ Wehner, \emph{The second laws of quantum thermodynamics},
Proc.\ Natl.\ Acad.\ Sci.\ USA \textbf{112}(11), 3275--3279 (2015).

\bibitem{Nielsen}
M.\ A.\ Nielsen, \emph{An introduction to majorization and its applications to quantum mechanics}, preprint, available at \url{http://michaelnielsen.org/blog/talks/2002/maj/book.ps}.

\bibitem{JonathanPlenio}
D.\ Jonathan and M.\ B.\ Plenio, \emph{Entanglement-Assisted Local Manipulation of Pure Quantum States}, Phys.\ Rev.\ Lett.\ \textbf{83}, 3566--3569 (1999).

\bibitem{Klimesh}
M.\ Klimesh, \emph{Inequalities that Collectively Completely Characterize the Catalytic Majorization Relation}, arXiv:0709.3680.

\bibitem{Turgut}
S.\ Turgut, \emph{Necessary and Sufficient Conditions for the Trumping Relation}, J.\ Phys.\ A: Math.\ Theor.\ \textbf{40}, 12185--12212 (2007).

\bibitem{Coecke}
B.\ Coecke, T.\ Fritz, and R.\ W.\ Spekkens, \emph{A mathematical theory of resources}, arXiv:1409.5531.

\bibitem{Aczel}
J.\ Acz\'el, B.\ Forte, and C.\ T.\ Ng, \emph{Why the Shannon and Hartley Entropies Are `Natural'}, Advances in Applied Probability \textbf{6}(1), 131--146 (1974).

\bibitem{Lostaglio}
M.\ Lostaglio, M.\ P.\ M\"uller, and M.\ Pastena, \emph{Stochastic Independence as a Resource in Small-Scale Thermodynamics}, Phys.\ Rev.\ Lett.\ \textbf{115}, 150402 (2015).

\bibitem{Fritz}
T.\ Fritz, \emph{Resource convertibility and ordered commutative monoids}, Math.\ Struct.\ Comput.\ Sci., FirstView, 1--89 (2016).

\bibitem{Faist}
P.\ Faist, F.\ Dupuis, J.\ Oppenheim, and R.\ Renner, \emph{The Minimal Work Cost of Information Processing}, Nat.\ Comm.\ \textbf{6}, 7669 (2015).

\bibitem{Burg}
J.\ P.\ Burg, \emph{Maximum entropy spectral analysis}, in Proc.\ 37th Meet.\ Society of Exploration Geophysicists, 1967. Reprinted in ``Modern Spectrum Analysis'', D.\ G.\ Childers, ed., New York, IEEE Press, 1978, pp.\ 34-41.

\bibitem{Gour}
G.\ Gour, M.\ P.\ M\"uller, V.\ Narasimhachar, R.\ W.\ Spekkens, and N.\ Yunger Halpern, \emph{The resource theory of informational
nonequilibrium in thermodynamics}, Physics Reports \textbf{583}, 1--58 (2015).

\bibitem{AczelDaroczy}
J.\ Acz\'el and Z.\ Dar\'oszy, \emph{On Measures of Information and Their Characterizations}, Academic Press, New York, 1975.

\bibitem{Csiszar}
I.\ Csisz\'ar, \emph{Axiomatic Characterizations of Information Measures}, Entropy \textbf{10}, 261--273 (2008).

\bibitem{Chakrabarti}
C.\ G.\ Chakrabarti and I.\ Chakrabarty, \emph{Shannon entropy: axiomatic characterization and application}, Internat.\ J.\ Math.\ Math.\ Sci.\ \textbf{2005}(17), 2847--2854 (2005).

\bibitem{Baez}
J.\ C.\ Baez, T.\ Fritz, and T.\ Leinster, \emph{A Characterization of Entropy in Terms of Information Loss}, Entropy \textbf{13}(11), 1945--1957 (2011).

\bibitem{Harremoes}
T.\ van Erven and P.\ Harremo\"es, \emph{R\'enyi Divergence and Kullback-Leibler Divergence}, IEEE Trans.\ Inf.\ Th.\ \textbf{60}(7), 3797--3820 (2014).

\bibitem{Linden}
N.\ Linden, M.\ Mosonyi, and A.\ Winter, \emph{The structure of R\'enyi entropic inequalities}, Proc.\ R.\ Soc.\ A \textbf{469}(2158), 20120737 (2013).

\bibitem{CoverThomas} T.\ M.\ Cover and J.\ A.\ Thomas, \emph{Elements of information theory}, Wiley \& Sons, New Jersey, 2006.

\bibitem{HHO}
M.\ Horodecki, P.\ Horodecki, and J.\ Oppenheim, \emph{Reversible transformations from pure to mixed states and the unique
measure of information}, Phys.\ Rev.\ A \textbf{67}, 062104 (2003).

\bibitem{Mate}
A.\ M\'at\'e, \emph{A new proof of a theorem of P.\ Erd\"os}, Proceedings of the AMS \textbf{18}(1), 159--162 (1967).

\bibitem{OHHH}
J.\ Oppenheim, M.\ Horodecki, P.\ Horodecki, and R.\ Horodecki, \emph{Thermodynamical Approach to Quantifying Quantum Correlations}, Phys.\ Rev.\ Lett.\ \textbf{89},
180402 (2002).

\bibitem{vanDam}
W.\ van Dam and P.\ Hayden, \emph{Universal entanglement transformations without communication}, Phys.\ Rev\ A \textbf{67}, 060302 (2003).

\end{thebibliography}
\end{document}